# Surface Termination Dependent Quasiparticle Scattering Interference and Magneto-transport Study on ZrSiS

Chih-Chuan Su[1§], Chi-Sheng Li[1§], Tzu-Cheng Wang[2§], Syu-You Guan[1,2], Raman Sankar[1,3], Fangcheng Chou[3], Chia-Seng Chang[1,2], Wei-Li Lee[1*], Guang-Yu Guo[2,4*], Tien-Ming Chuang[1*]

[1] Institute of Physics, Academia Sinica, Nankang, Taipei 11529, Taiwan

[2] Department of Physics and Center for Theoretical Physics, National Taiwan University, Taipei 10617, Taiwan

[3] Center for Condensed Matter Sciences, National Taiwan University, Taipei 10617, Taiwan

[4] Physics Division, National Center for Theoretical Sciences, Hsinchu 30013, Taiwan

[§] These authors contributed equally to this work

[*] To whom correspondence should be addressed.

E-mail: wlee@phys.sinica.edu.tw, gyguo@phys.ntu.edu.tw, and chuangtm@sinica.edu.tw

## Abstract

Dirac nodal line semimetals represent a new state of quantum matters in which the electronic bands touch to form a closed loop with linear dispersion. Here, we report a combined study on ZrSiS by density functional theory calculation, scanning tunnelling microscope (STM) and magneto-transport measurements. Our STM measurements reveal the spectroscopic signatures of a diamond-shaped Dirac bulk band and a surface band on two types of cleaved surfaces as well as a spin-polarized surface band at $\bar{\Gamma}$ at E~0.6eV on S-surface, consistent with our band calculation. Furthermore, we find the surface termination does not affect the surface spectral weight from the Dirac bulk bands but greatly affect the surface bands due to the change in the surface orbital composition. From our magneto-transport measurements, the primary Shubnikov-de-Haas frequency is identified to stem from the hole-type quasi-two-dimensional Fermi surface between $\Gamma$ and X. The extracted non-orbital magnetoresistance (MR) contribution $D(\theta, H)$ yields a nearly H-linear dependence, which is attributed to the intrinsic MR in ZrSiS. Our results demonstrate the unique Dirac line nodes phase and the dominating role of Zr-$d$ orbital on the electronic structure in ZrSiS and the related compounds.

# Introduction

The study on Dirac-like systems has progressed greatly in the past decade. While two-dimensional (2D) Dirac fermions were found in graphene [1] and on the surfaces of topological insulators [2], three-dimensional (3D) Dirac fermions that are protected by inversion symmetry and time reversal symmetry were demonstrated experimentally in $Na_3Bi$ [3] and $Cd_3As_2$ [4]. Weyl fermions that stem from Dirac fermions due to the broken inversion symmetry or broken time reversal symmetry were also observed e.g. in TaAs [5]. Furthermore, it has been predicted that Dirac nodal line semimetal phase exists when the non-degenerate conduction and valence bands cross each other to form a closed loop [6]. Experimentally, several Dirac nodal line semimetals, within which their linear bands crossings connect to form continuous lines or arcs, were observed in $PbTaSe_2$ [7], $PtSn_4$ [8] and ZrSiS family [9-13]. However, it has been revealed that both $PbTaSe_2$ [7, 14] and $PtSn_4$ [8] exhibit the complex band structure, making them difficult to study the Dirac nodal bands separately. In contrast, the electronic band structure of ZrSiS is relatively simple. Indeed, recent ARPES studies on ZrSiS revealed the existence of non-symmorphic symmetry protected line nodes [9-13], which was first predicted by Kane and Young in 2015 [15]. In addition, there also exist a Dirac surface band at $\overline{X}$ and a diamond-shaped Dirac band at $\overline{\Gamma}$ from ARPES [9-13, 16] and scanning tunnelling microscope (STM) study [17, 18]. Both Dirac bands are also shown to disperse linearly up to few hundred meV even above $E_F$ by quasiparticle scattering interference (QPI) imaging [17, 18], from which ZrSiS exhibits strongly band-selective scattering depending on impurity lattice site [17]. Moreover, transport measurements revealed extremely large, non-saturated and anisotropic magnetoresistance (MR) in ZrSiS [16, 19-21]. Several Shubnikov-de-Haas (SdH) oscillations measurements reported a major peak at ~240 T, which is attributed to the quasi-2D Fermi surface [16, 19, 21-25] while a peak at ~600 T observed by thermoelectric power [23] and high field SdH [24] measurement is assigned to a 3D Fermi surface. Recently, frequency independent flat optical conductivity [25], the rapid relaxation rate after photoexcitation [26] and Ag tip induced superconductivity [27] are also reported in ZrSiS. Thus, these fascinating phenomena demonstrated that the ZrSiS family is a great system for studying the Dirac nodal-lines [11, 12, 28, 29]. However, the experimental investigations on the detail electronic band structure above $E_F$ and the orbital contribution are still inadequate. In addition, a proper model to extract the non-orbital MR contribution is needed to reveal the intrinsic MR contribution due to the Dirac fermions.

In this paper, we investigate the surface and bulk electronic structures in Dirac nodal line semimetal, ZrSiS, by density function theory (DFT) calculations, STM and magneto-transport measurements. Our bulk calculation reveals that the Fermi surface of ZrSiS primarily consists of a 3D diamond-shaped Dirac band at Γ and a quasi-2D tubular band between Γ and X. The slab calculation shows a surface band exists at X̄ with linear dispersion over several hundred meV. A fully spin-polarized surface state at $E \sim 0.6$ eV is found to connect the diamond-shaped Dirac band and the bulk conduction band and is completely separated from the bulk bands. Our STM measurements reveal both S- and Si-surfaces and we deduce the observed atoms to be Zr, independent of the surface terminations. Our data from QPI imaging support the band structure from our DFT calculation. Our further analysis suggests that different surface terminations have negligible effect on the surface spectral weight from the Dirac bulk bands in ZrSiS but that of the surface bands can be greatly reduced due to the change in orbital composition. The enhanced surface spectral weight from the diamond-shaped band produces a stronger QPI pattern than that from the surface band. Our finding suggests an unusual scattering selection rule and highlights the important role of Zr-$d$ orbitals. On the other hand, multiple SdH frequencies are clearly identified from the magneto-transport measurements. A model is proposed and applied on the angular dependent MR data to separate the dominant orbital MR contribution and the non-orbital MR that is likely associated with the Dirac bands. Our results thus demonstrate that ZrSiS and its related compounds are a valuable platform for exploring Dirac line nodes physics and their future applications.

## Band structure calculation

Our ZrSiS single crystals are grown by two-step chemical vapour transport process as described in detail elsewhere [30] and have the PbFCl-type crystal structure in tetragonal space group *P4/nmm*, as showed in figure 1(a). The lattice constants, determined by x-ray diffraction, are a = 3.544 Å and c = 8.055 Å. In ZrSiS, square $Zr_2S_2$ layers are alternatively stacked along the c-axis with double layers of Si atoms. We calculate the electronic structure based on DFT theory. The bulk band structure without and with SOC is shown in figures 1(b) and 1(c), respectively. Several Dirac bands can be found in the bulk band structure without SOC (circled in red and blue). The Dirac band crossings also form two one-dimensional (1D) line nodes between R and X at E ~ -0.45 eV to -0.67 eV and also between A and M at $E \sim -2.2$ eV. When the SOC is taken into account in the calculation, the Dirac bands near the Fermi level (marked as red circles in figures 1(b) and 1(c)) open a small gap (~37 meV) but their dispersion remains linear as illustrated in figure 1(c). In contrast, the 1D line nodes (marked as blue circles in figures 1(b) and 1(c)) in the occupied states remain degenerate due to the protection from non-symmorphic symmetry [9-13, 17, 31]. These Dirac points are connected to form a 3D diamond-shaped Fermi surface at Γ. Moreover, quasi-2D tubular

Fermi surface is also found along ΓX as shown in figure 1(f). These results are consistent with recent ARPES [9-13, 16, 31] and SdH [16, 19, 21, 22, 24, 25] measurements on ZrSiS. To better compare the theoretical calculation with STM and ARPES measurements, we also carry out a slab calculation and the resultant band structure with the surface component is shown in figure 1(d). By comparing the electronic structure of ZrSiS from the bulk and slab calculations, we find the surface states with linear dispersion around $\bar{X}$ at $E_F$, and this is consistent with previous ARPES [9-13] and STM [17] experiments. Interestingly, we also find that the SOC opens a gap of ~100 meV (orange circles in figure 1(b) and (c)) on the Dirac crossings around $E \sim 0.6$ eV and a surface state emerges to connect the top of inner Dirac band and the bottom of conduction bands (orange arrow along $\overline{\Gamma M}$ in figure 1(d)). Our DFT calculation also shows that this surface state is fully spin-polarized and the spins are all aligned in-plane clockwise as showed in figure 1(e). The surface state is still protected by time reversal symmetry but not by topology since our calculation shows ZrSiS is topologically trivial. From the calculated band structure, the linear dispersion of all Dirac bands in ZrSiS extends over a wide energy range of several hundred meV, which has a great advantage over other Dirac semimetals with linear dispersion of much smaller energy range.

### STM experiments

We first investigate the surface electronic structure with STM measurements, which are carried out in a homemade low temperature ultra-high vacuum STM system at $T = 4.5$ K with electrochemically etched tungsten tips [33]. The STM tips are cleaned and reshaped first by field-emission on the gold surface. The ZrSiS single crystal is *in situ* cleaved at the sample stage of the STM head at around $T = 5$ K right before STM measurements. Due to the nature of a layered crystal structure with a glide plane between S layers, it is generally assumed that the cleavage should occur naturally here and yield large flat and charge neutral S-surface. Interestingly, we find that ZrSiS can also be cleaved to reveal Si-surface. Figure 2(a) shows a $80 \times 80$ nm$^2$ topographic image of ZrSiS surface with three large terraces. The linecut shows the step height of 8.46 and 3.61 Å, from which we can identify two S-surfaces and a Si-surface as illustrated in figure 2(a) and appendix A3. The high resolution topographic images acquired on S- and Si-surface are shown in figures 2(b) and 2(c), respectively. Surprisingly, both terminations show the $1 \times 1$ atomic structure with the same lattice constant of $3.54 \pm 0.11$ Å and the same orientation, which matches that of S-layer or Zr-layer. When the STM tip is moved closer to the surface by taking the topographic image with higher current, each surface shows different features. Additional atoms become visible at the centre of the square lattice on the S-surface as shown in figure 2(d) while the atoms on the Si-surface become less clear due to hybridization between atoms along a-axis in figure 2(e). Thus, we deduce that the Zr atoms are observed and the surface atoms are invisible from STM in figures 2(b) and (c). The typical differential tunnelling spectra measured on the clean regions of both surfaces show almost identical characteristic semimetallic spectra as illustrated in figure 2(c). The local density of states (LDOS) near Fermi level is the lowest due to Dirac crossing in the band structure but not zero. The measured tunnelling spectra appear to be in excellent agreement with our orbital-projected density of states (PDOS) calculation, as displayed in figure 2(f), in which Zr-$d$ orbital dominates the LDOS near $E_F$ over S-$p$ and Si-$p$ orbitals on the surface. This explains why it is easier to observe Zr atoms than S or Si atoms in STM topographic images and confirms the previous impurities study of ZrSiS by STM and DFT study [17]. The simulated STM images also support this view [17]. A dip exists at $E \sim 0.6$ eV in PDOS calculation and is also observed in tunnelling spectrum at $E \sim 0.74$ eV, which could be related to the change in energy dispersion as indicated by black and orange arrows in figure 1(d). This dip feature will also allow us to better compare the experimental data with DFT results at the high energy in the following section.

To further investigate the dependence of the electronic structure on differently terminated surfaces, we utilize Fourier transform-scanning tunnelling spectroscopy (FT-STS), which can relate the real space characteristic wavevectors resulting from QPI to the band structure in momentum space. First, energy resolved differential tunnelling conductance maps, dI/dV($\mathbf{r}$, E = eV) are taken in a field of view (FOV) of $60 \times 60$nm$^2$ on S- and Si-surfaces as shown in figure 3(a)-(e) and (k)-(o), respectively. The standing waves due to QPI are visible around the defects and impurities and show very similar patterns at the same energy. The main difference is that many impurities on the Si-surface do not contribute to any QPI (possibly S clusters, as shown in appendix A4) and the QPI signals diminish at E < ~125 meV. We then take FT of the dI/dV($\mathbf{r}$, E) maps and obtain the corresponding dI/dV($\mathbf{q}$, E) maps on S- and Si-surfaces, as shown in figure 3(f)-(j) and (p)-(t), respectively (the movies also available in supplementary materials, are available online at stacks.iop.org/NJP/20/ 103025/mmedia). Three pronounced dispersive C$_4$-symmetric $\mathbf{q}$-vectors ($\mathbf{q_1}$, $\mathbf{q_2}$ and $\mathbf{q_3}$) can be observed in the dI/dV($\mathbf{q}$, E) maps (figure 3). After analysing the direction, wavelength and energy dispersion of $\mathbf{q}$-vectors and comparing with the calculated band structure, we can identify $\mathbf{q_1}$ as the intra-pocket scattering between the diamond-shaped band at $\bar{\Gamma}$, $\mathbf{q_2}$ as the intra-pocket scattering of the surface band at $\bar{X}$ and $\mathbf{q_3}$ as the inter-pocket scattering between the surface bands at $\bar{X}$ as shown in figure 4(a). Our $\mathbf{q}$-vectors assignments are consistent with previous QPI studies on ZrSiS [17, 18]. However, on the Si-surface, all $\mathbf{q}$-vectors become weaker with decreasing energy and the QPI from the surface bands ($\mathbf{q_2}$ and $\mathbf{q_3}$) diminished at E < ~125 meV. This is mainly because the disorder from the large amount of impurities on the Si-surface suppresses the QPI. In addition, the surface band near $\overline{X}$ has significant S-$p$ orbital contribution and the dispersion towards $\overline{\Gamma}$ and $\overline{M}$ is linked to a change in its Zr-$d$ orbitals [31], which reduces the spectral weight on the surface bands and in term, weakens the QPI near $\overline{X}$ on the Si-surface. To quantify the QPI data, we then extract $q_1$ and $q_2$ from the dI/dV($q$, E) maps taken on both surfaces (see figures 4(b) and (c)), in which both $q$-vectors exhibit the identical energy dispersion on both surfaces within the overlapped energy range and agree well with our calculated band structure. The small difference between our QPI and DFT data at the high energy can be accounted if we take into account an energy-shift by comparing the dip at E ~ 0.74 eV in tunnelling spectrum with PDOS. Figures 4(b) and (c) show that the diamond-shaped Dirac bulk bands at $\overline{\Gamma}$ exhibit the linear dispersion from E ~ -0.1 eV to E ~ +0.5 eV and the surface bands at $\overline{X}$ from E ~ -0.1 eV to E ~ 0.8 eV. The derived Fermi velocity of the bulk Dirac band is ~$1.6\times10^6$ m/s and of the surface band is ~$8.2\times10^5$ m/s, both of which are comparable with that in graphene. Our results demonstrate that the surface termination does not affect the bulk Dirac bands in ZrSiS; however, the spectral weight of the surface bands is greatly reduced due to the change in orbital composition.

To verify the existence of the predicted spin-polarized surface at higher energy, we take dI/dV($r$, E) maps between 500 and 1000 meV with higher $q$-resolution on the same FOV of the S-surface (figure 3(a)) because it will be difficult to resolve smaller $q_1$ as shown in figures 5(a)-(c). The corresponding dI/dV($q$, E) maps are shown in figures 5(d)-(f), in which $q_1$ vanishes at E > ~660 meV and appears again at E > ~860 meV. This energy range coincides with the dip in tunnelling spectrum and PDOS in figure 2(f). Thus, we can conclude that the disappearance of $q_1$ is consistent with the time reversal symmetry protection from the predicted fully spin-polarized surface band in figure 1(d) and (e). We are unable to make the comparison with data from Si-surface because of the large background at $q$ = 0 due to the large amount of surface impurities. Although such a surface state is not protected by the non-symmorphic symmetry nor the band topology, its full in-plane spin polarization with time reversal protection may be useful in a related compound that has the Fermi level within its proximity.

**Magneto-transport measurements**

To further study the bulk electronic structure and transport properties, we carry out magneto-transport measurements by standard four-probe geometry. The resistivity equals about 9 μΩcm at room temperature and gradually decreases with decreasing temperature, showing a metallic behavior (figure 6(a)). Below 20 K, the resistivity is nearly independent of the temperature with a residual resistivity of ~

0.23 μΩcm, giving a residual resistivity ratio (RRR $\equiv \rho_{300K}/\rho_{10K}$) of ~ 38.8. Figure 6(b) shows the magnetoresistance (MR$\equiv [\rho(H)/\rho(0)] - 1 = \Delta\rho/\rho$) data at $T = 2$ K and at several different $\theta$ values with a magnetic field up to 15 Tesla, where $\theta$ is defined as the angle between the magnetic field and the normal direction of the sample surface as illustrated in the upper inset of figure 6(b). Basically, it exhibits a positive MR, but its magnitude does not vary monotonically with $\theta$. The largest MR turns out to appear at about $\theta = 30°$, giving a $\Delta\rho/\rho \approx 173$ at 15 Tesla. As $\theta$ further increases, the MR gradually drops to about $\Delta\rho/\rho = 2.7$ at $\theta = 90°$ and 15 Tesla. The MR data at $\theta = 90°$ in figure 6(b) are multiplied by 10 for visibility. We note that, in lower fields, MR curves show crossover from concave curvature to convex curvature at $\theta = 90°$, where practical field linear MR is observed for $\mu_0H \leq 1$ Tesla at $\theta = 90°$. The MR data are replotted in polar figure shown in figure 6(c) at several different fields, which shows a butterfly-like pattern with two-fold symmetry.

As shown in figure 6(b), apparent SdH oscillations are observed as the field becomes larger than 3 Tesla. The corresponding fast Fourier transformation (FFT) spectra at different $\theta$ values ranging from 0° to 90° are shown in figure 7(a), where the curves are shifted for clarity and the curves for $\theta = 75°$ and 90° are further amplified by 8-fold and 10-fold, respectively. At $\theta = 0°$, three peaks of $F_1$, $F_2$ and $F_3$ can be clearly identified from the FFT spectrum as denoted in figure 7(a). The major SdH frequency $F_1$ equals 240.3 Tesla. $F_2$ is nearly twice of $F_1$ and likely to be the second harmonic contribution. $F_3$, on the other hand, gives a small SdH frequency of about 18.4 Tesla. The corresponding d$\rho$/dH data versus $1/\mu_0H_\perp \equiv 1/\mu_0Hcos\theta$ at $\theta = 0°$, 15°, 30°, 45° and 60° are shown in figure 7(b), where beating patterns are observed and justified the multiple frequencies in the FFT spectra. The SdH oscillation period for $F_1$ is found to scale as $1/\mu_0H_\perp$, indicating a 2D-like band structure that is likely deriving from the hole-type quasi-2D tubular Fermi surface shown in figure 1(f). This 2D-like feature is further demonstrated by the dashed line in figure 7(a) and also the excellent linear fitting to the $F_1$- $1/cos\theta$ curve in the lower-left inset in figure 7(b), which is in big contrast to the weak θ dependence of $F_3$. For magnetic fields along c axis ($\theta = 0°$), the hole-type quasi-2D tubular Fermi surface gives an extremal Fermi surface area of $A_F = 0.023084$ Å$^2$, which corresponds to a SdH frequency of about 241.8 Tesla according to the Onsager relation (F=(h/4π$^2$e)$A_F$). This turns out to agree well with the observed major SdH frequency of $F_1$= 240.3 Tesla. We note that, as $\theta \geq 30°$, the major frequency $F_1$ appears to get broader and then split into multiple peaks, where the peak separation grows larger as $\theta$ increases. This behaviour may be related to the fine structures in the 2D tubular Fermi surface shown as the blue area in figure 1(f). The observed major SdH frequency

F1 agrees with several previous reports [24, 29, 34], but we do note that the 3D diamond-shaped band (the green area in figure 1(f)) was not seen in our SdH measurements, which is likely due to its reported unusual mass enhancement [29, 35] and our limited field strength.

The presence of multiple bands conduction in ZrSiS is evident from the non-linear field dependence of Hall resistivity (not shown here), and multiple SdH frequencies. In addition, the calculated band structure also suggests the existence of both electron and hole bands at the Fermi level. To find out the intrinsic MR from the bulk bands, we carry out the following analysis. According to the two-band model, the corresponding orbital MR contribution can be expressed as

$$(\Delta\rho/\rho)_{orb} = \frac{n_1\mu_1 n_2\mu_2(\mu_2-\mu_1)(\mu_0 H)^2}{(n_1\mu_1+n_2\mu_2)^2+(n_1+n_2)^2\mu_1^2\mu_2^2(\mu_0 H)^2} \ , \qquad (1)$$

where $\mu_{1(2)}$ and $n_{1(2)}$ are the mobility and density, respectively, of the first (second) band. $(\Delta\rho/\rho)_{orb}$ is only sensitive to the perpendicular component of the field, and it scales as $(\mu_0 H_\perp)^2$ in lower fields. Based on the Matthiessen's rule, the total MR can thus be expressed as

$$(\Delta\rho/\rho)(\theta, H) = C(\theta)(\mu_0 H cos\theta)^2 + D(\theta, H), \qquad (2)$$

where $(\Delta\rho/\rho)_{orb} = C(\theta) \ (\mu_0 H cos\theta)^2$, $C(\theta)$ is a $\theta$-dependent prefactor and $D(\theta, H)$ term represents the non-orbital contribution to the MR. Assuming that $D(\theta, H)$ term grows with field at an exponent less than 2, the value of $(\Delta\rho/\rho)(\theta, H)/ \ (\mu_0 H cos\theta)^2$ should gradually approach $C(\theta)$ in higher fields, where $D(\theta, H)/ \ (\mu_0 H cos\theta)^2$ term becomes vanishingly small. Figure 8(a) shows the $(\Delta\rho/\rho)(\theta, H)/ \ (\mu_0 H cos\theta)^2$ versus $\theta$ curves at different fields ranging from 0.3 Tesla to 15 Tesla, where the curve indeed gradually approaches a field independent function that is shown as a thick red line for $\mu_0 H = 15$ Tesla. By taking the thick red line as $C(\theta)$ and using equation (2), we can extract the non-orbital part MR contribution $D(\theta, H)$. The corresponding $D(\theta, H)$ versus $H$ curves at different $\theta$ values are shown in figure 8(b). Surprisingly, the $D(\theta, H)$ is practically linear in $H$ field for $\mu_0 H \ \leq 4$ Tesla with a drop in the slope by about 4-fold as $\theta$ value goes from $0^o$ to $75^o$. Figure 8 (c) replots the $D(\theta, H)$ versus $H$ in a log scale, where the orange dashed line is the linear fit to the data. We remark that $H$-linear dependency in $D(\theta, H)$ persists down to a very weak field of about 0.3 Tesla, and the extracted exponent $\alpha$, defined as $D(\theta, H) \propto H^{\alpha}$, as a function of $\theta$ is shown in figure 8(d). The exponent $\alpha$ is found to be close to 1.0 and nearly independent of the $\theta$ values, which also justifies the assumption we made earlier ($\alpha < 2$) for the analysis using Eq. (2). We do note the failure of $H$-linearity in $D(\theta, H)$ for $\mu_0 H > 4$ Tesla as can be seen in figure 8(c), where we attribute such deviation from $H$-linearity to the additional contribution from quantum oscillations in the quantum regime. Therefore, the intrinsic MR from the bulk band effect is likely to be best revealed in the low field regime as shown in figure 8(b).

The $H$-linear MR phenomenon has been observed in topological Dirac and Weyl semimetals, such as Dirac semimetal $Cd_3As_2$ [36] and Weyl semimetal NbP [37, 38]. Combining the observation of linear dispersion band from QPI analysis and also special Dirac line node in the calculated band in ZrSiS, the observed $H$-linear $D(\theta, H)$ may provide further support for the existence of the Dirac-like bulk band from the magneto-transport measurements. Nevertheless, the correct theoretical description for the $H$-linear MR due to Dirac-like band remains a debatable issue [38-41]. On the other hand, it is also suggested that similar $H$-linear MR effect can be observed in a narrow band gap semiconductor with inhomogeneous conductivity [40, 41]. As shown in figure 8(b), $D(\theta, H)$ appears to be at maximum at $\theta = 0^o$, and the slope d$D(\theta, H)$/dH seems to roughly scales as cos $\theta$, which may rule out the possible Zeeman-like spin spitting effects on Dirac electrons[38]. We also note that the $H$-linear in $D(\theta, H)$ remains valid down to weak field regime, suggesting the quantum MR due to the filling of single Landau level may not apply either [39]. Further investigation is needed to elucidate the intrinsic mechanism for the $H$-linear MR effect in ZrSiS.

## Summary

In summary, we have reported a combined study on the bulk and surface electronic structures of ZrSiS by using DFT, STM and magneto-transport measurements. Our band calculations show that ZrSiS exhibits 3D Dirac line nodes protected by non-symmorphic symmetry, Dirac surface bands at $\bar{X}$, a fully spin-polarized surface band at $\bar{\Gamma}$ at E ~ 0.6 eV. ZrSiS has a 3D electron-like diamond-shaped Fermi surface and a quasi-2D hole-like tubular Fermi surface. Our results from QPI imaging and SdH measurements show overall excellent agreement with the calculated band structures. Moreover, both FSs have strong Zr-$d$ orbital contribution and high carrier mobility, which can contribute the butterfly-shaped MR. We analyse our angular dependent MR data with a two-band model and deduce the H-linear non-orbital part of MR at H < 4 Tesla is most likely intrinsic. Our surface termination dependent topographs and QPI imaging demonstrate that the Zr-$d$ orbitals contribution to the surface spectral weight of the Dirac diamond-shaped bulk band is robust against the surface modification whereas the orbital composition of the Dirac surface bands depends on the type of surface atoms. Our results support the importance of Zr-$d$ orbitals on the electronic structures and the unusual Dirac line nodes phase in ZrSiS, which can also be generalized to the related compounds. Thus, the ZrSiS family provides a rich platform to explore the effect of non-symmorphic symmetry and orbital degree of freedom on their electronic structures.

## Acknowledgements

We thank Andreas W Rost, Hsin Lin, Peng-Jen Chen, Tay-Rong Chang, Minn-Tsong Lin, Christopher J. Butler, Horng-Tay Jeng and Sungkit Yip for helpful discussions. This work is supported by Academia Sinica (AS-iMATE-107-12), Ministry of Science and Technology, Taiwan and National Taiwan University. T-MC and G-YG are grateful for the generous support from Golden Jade Fellowship of Kenda Foundation.

**Figures**

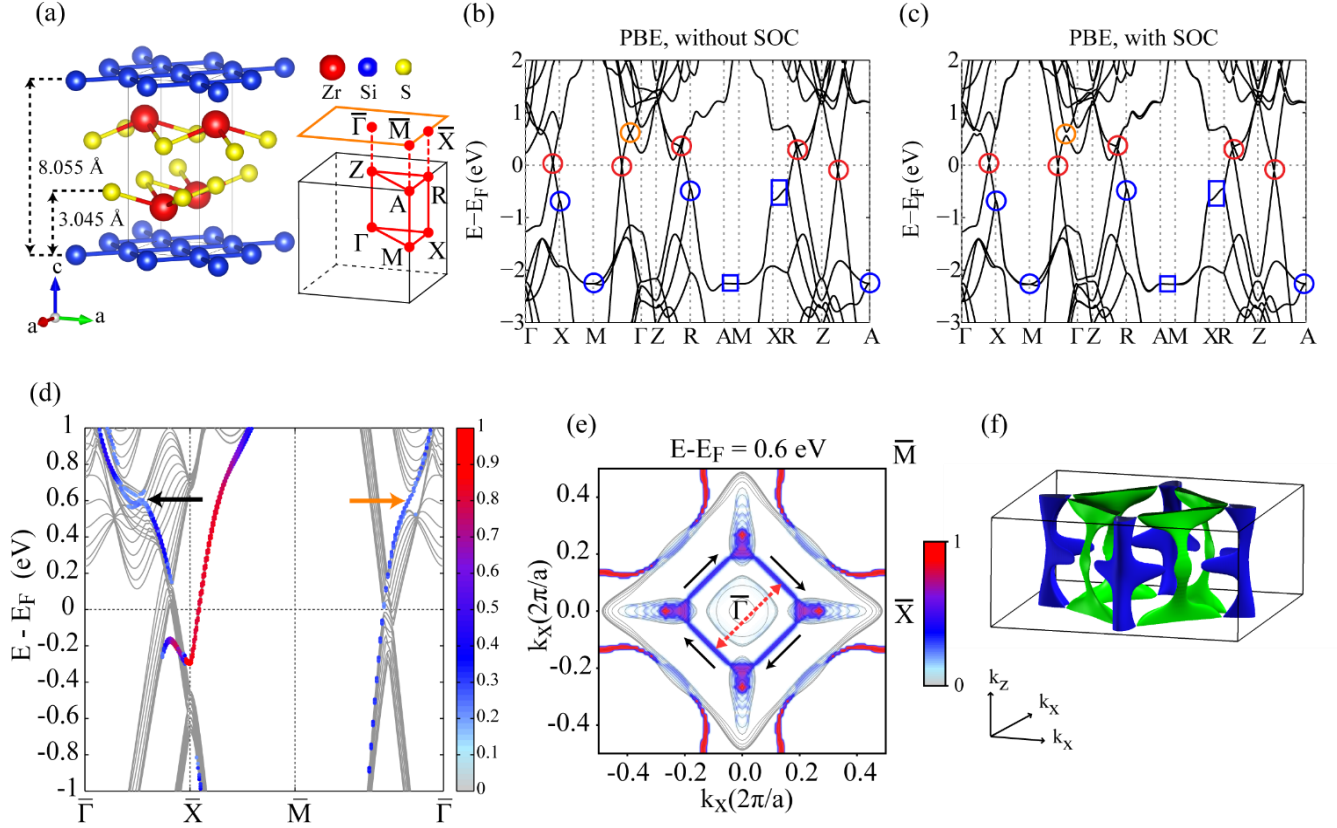

**Figure 1.** Crystal structure and band structure of ZrSiS. (a) Crystal structure of ZrSiS, visualized using VESTA [32] (left) and its corresponding Brillouin zone (right) of bulk (black) and S-surface (orange). Bulk band structure without (b) and with (c) SOC. The Dirac crossing points are marked by blue and red circles. SOC opens a gap on the nodes marked in red whereas the nodes marked in blue are protected by non-symmorphic symmetry and form the Dirac line nodes in the regions marked in blue rectangles. (d) Surface band structure of S-surface from a 9-layers slab calculation. The colorbar represents the ratio of surface components of the band. (e) Slab-calculated constant energy contour of ZrSiS at $E = 0.6$ eV. The colorbar shows the ratio of surface components as in (d). The diamond-shaped surface band at $\bar{\Gamma}$ (indicated by the orange arrow in (d)) is separated from the bulk bands. Our calculation also shows its helical spin texture is fully in-plane and rotates in clockwise as indicated as black arrows. The red dashed arrow indicated the forbidden back scattering $\boldsymbol{q}$-vector. (f) The 3D Fermi surface from the bulk band structure in (c). The green and the blue areas represent the electron-like Dirac diamond-shaped band and the hole-like quasi-2D tubular band, respectively.

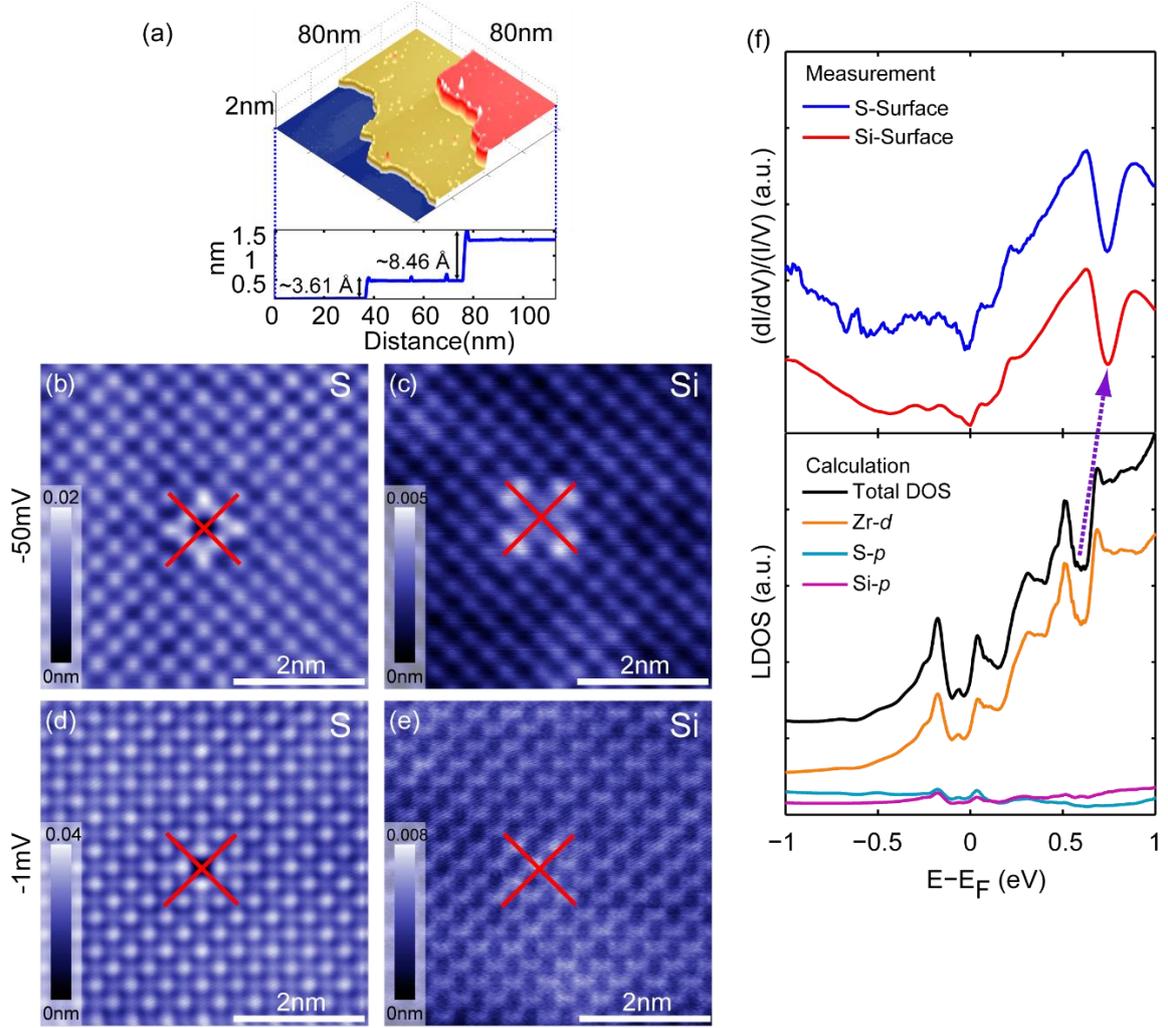

**Figure 2.** STM topographic images and tunnelling spectra. (a) Topographic image in a $80 \times 80$ nm$^2$ FOV shows multiple steps and terraces. The height profile along the diagonal direction in (a) is shown in the bottom, from which we deduce the blue area is Si-surface and the other two terraces are S-surfaces. (b) and (c) The topographic images taken on S- and Si-surface, respectively with bias = -50 mV and I = 0.5 nA show the identical atomic structure. (d) and (e) The topographic images taken on S- and Si-surface with bias = -1 mV, I = 0.5 nA and 2 nA for (d) and (e), respectively. The red crosses indicate the impurities location as reference. (f) The measured tunnelling spectrum (top) and the calculated PDOS (bottom). The purple arrow indicates the corresponding dip feature, which we use to estimate the required energy-shift at higher energy between the measurement and calculation. The measured tunnelling spectrum on S-surface (blue) is shifted by 1 for clarity.

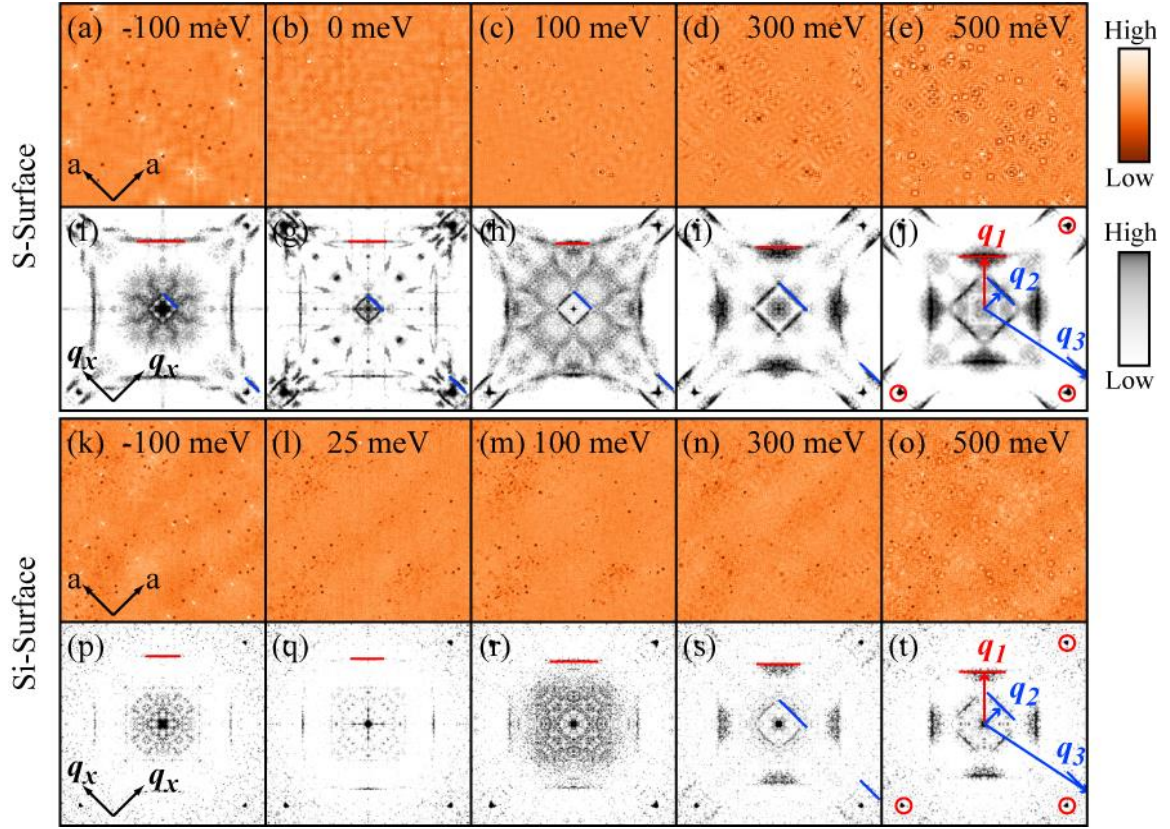

**Figure 3.** QPI imaging and Fourier analysis on ZrSiS surface. Normalized conductance dI/dV($\mathbf{r}$, E) maps at five energies taken on (a)-(e) S-surface and (k)-(o) Si-surface, respectively. All images are taken in the same $60 \times 60$ nm² FOV on the respective surface at T = 4.5 K as shown in appendix A4. The corresponding Fourier transformed images, dI/dV($\mathbf{q}$, E) on (f)-(j) S-surface and (p)-(t) Si-surface, respectively. The red circles indicate $2\pi/a$, where a is the lattice constant of Zr. The red line-segments and the red arrows indicate $\mathbf{q_1}$. The blue line-segments and the blue arrows indicate $\mathbf{q_2}$ and $\mathbf{q_3}$.

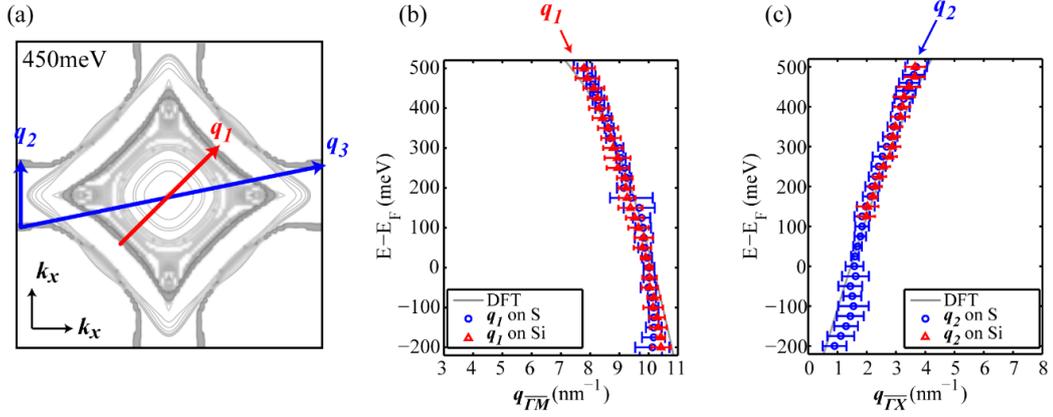

**Figure 4.** Energy dispersion of $\boldsymbol{q}$-vectors from QPI imaging. (a) The grey lines represent the calculated constant energy contour at E = 450 meV. Three arrows which represent three major $\boldsymbol{q}$-vectors observed in the QPI data and their relation to the band structure. (b) The energy dispersion of $\boldsymbol{q_1}$ along $\overline{\Gamma M}$. (c) The energy dispersion of $\boldsymbol{q_2}$ along $\overline{\Gamma X}$. The blue circles and the red triangles represent the data extracted from the measured dI/dV($\boldsymbol{q}$, E) maps taken on S- and Si-surface, respectively. The grey solid lines show the expected QPI dispersion from calculated band structure. The dispersion of $\boldsymbol{q_1}$ and $\boldsymbol{q_2}$ on both surfaces are identical but $\boldsymbol{q_2}$ on Si-surface diminishes below E ~ 125 meV.

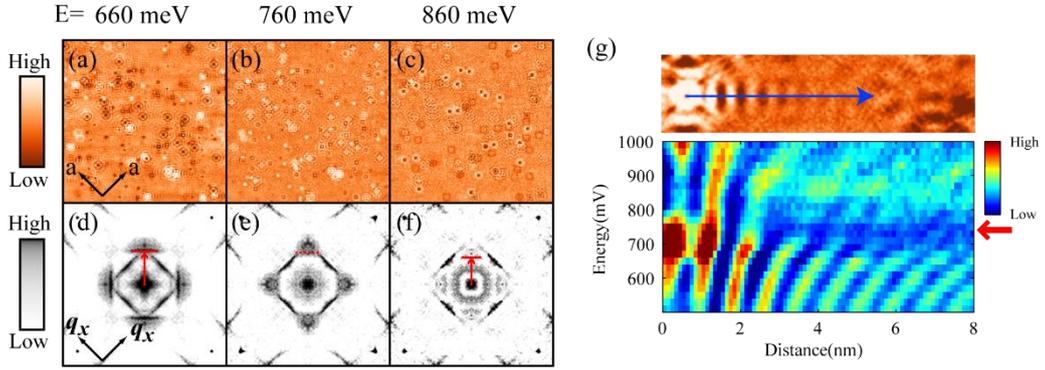

**Figure 5.** QPI signature of the fully spin-polarized surface band. (a)-(c) Normalized conductance dI/dV(**r**, E) maps taken on S-surface at three different energies. All images are taken in the same FOV at T = 4.5 K as shown in appendix A4. (d)-(f) The corresponding Fourier transformed images, dI/dV(**q**, E). The red line-segments and the red arrows representing **q₁** vanish between 660 and 860 meV. The red dashed line in (e) marks the expected position of **q₁** but it is absent in our data due to the forbidden back scattering as shown in figure 1(e). (g) The real space QPI wavevector as a function of energy taken around an impurity site that produces **q₁**. The line profile (bottom) is taken along the blue arrow in the topograph (top). The QPI signals also diminish in the same energy range as in (a)-(c), which is the low DOS region marked by the red arrow. These results support the predicted fully spin-polarized surface band and the intraband back scattering is forbidden by the time reversal symmetry.

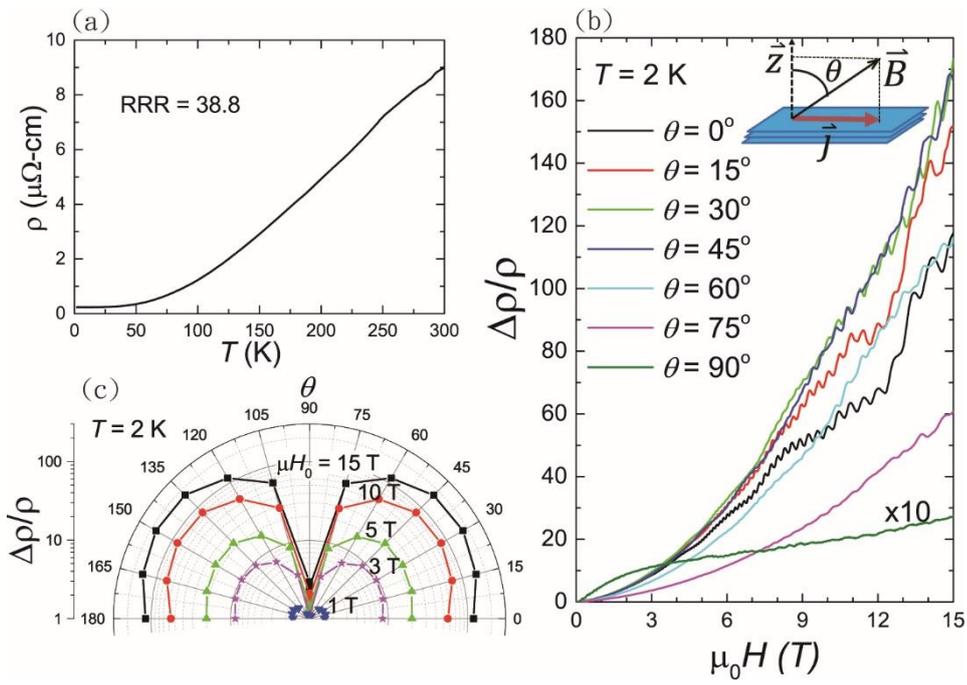

**Figure 6.** (a) The temperature dependence of the resistivity in ZrSiS. (b) MR at $T = 2$ K and various $\theta$ values ranging from 0º to 90º. The upper inset cartoon illustrates the angle definition for $\theta$. (c) The corresponding polar plot of MR at $T = 2$ K in various magnetic fields.

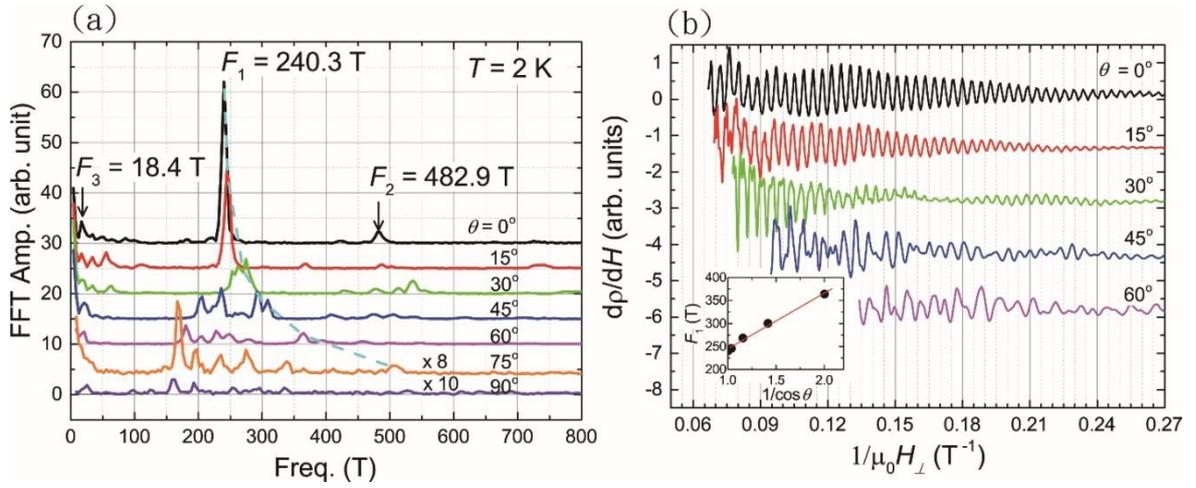

**Figure 7.** (a) FFT spectrum of the SdH oscillation at various $\theta$ values. Three SdH frequencies can be clearly identified and labeled as $F_1$, $F_2$, and $F_3$. (b) The derivative of resistivity $d\rho/dH$ as a function of $1/\mu_0 H_\perp$. The inset figure shows $F_1$ versus $1/cos\theta$, and the red line is the linear fit to the data.

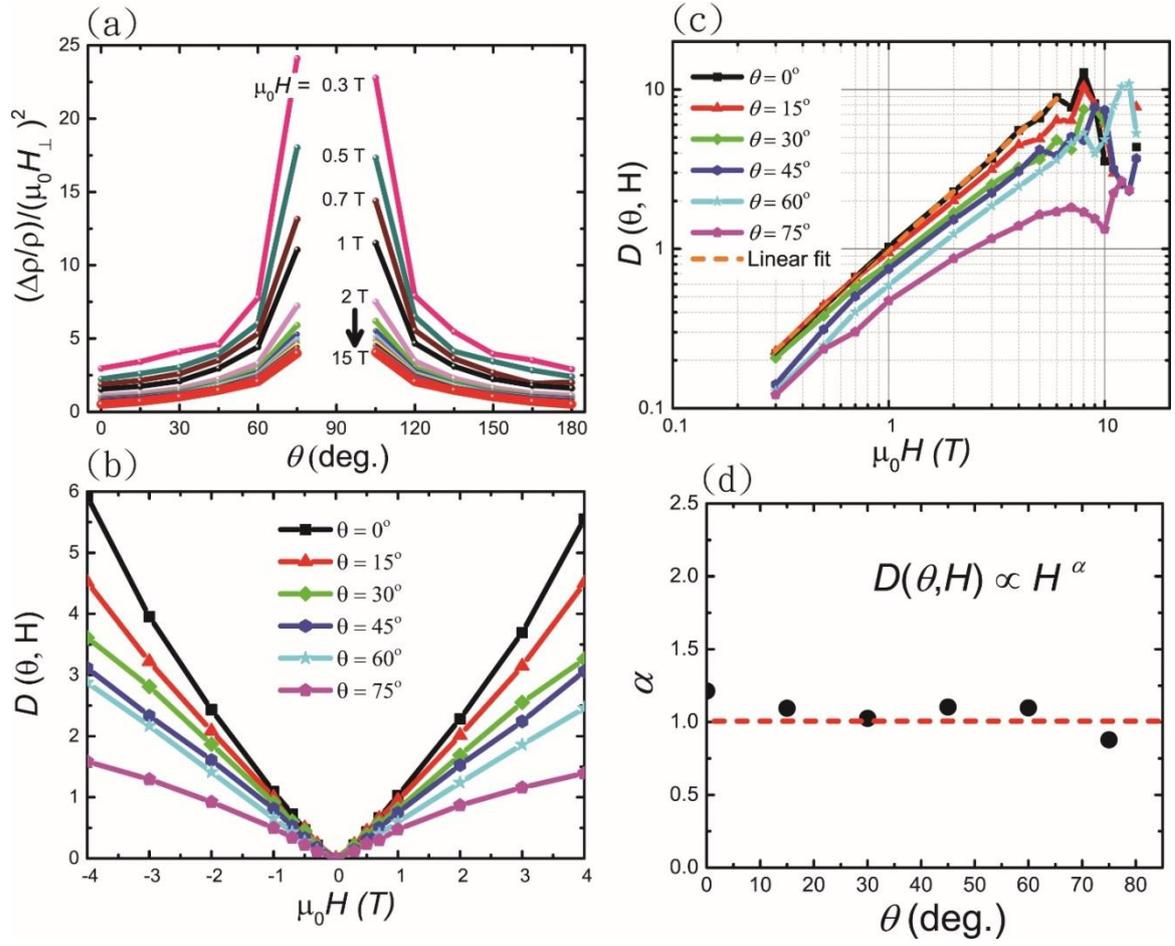

**Figure 8.** (a) $(\Delta\rho/\rho)/(\mu_0 H_\perp)^2$ vs $\theta$ at different magnetic fields up to 15 Tesla. (b) $D(\theta, H)$ as a function of $\mu_0 H$, and the corresponding log-scale plot is shown in (c). The orange line is a linear fit to the data. (d) The extracted exponent $\alpha$ as a function of $\theta$, where $\alpha$ is close to 1.0 and nearly independent of $\theta$.

## Appendix

### A1. Computational details

The electronic band structure calculations are carried out within the density functional theory (DFT) with the generalized gradient approximation (GGA) to the electron exchange-correlation potential [42]. The accurate full-potential projector-augmented wave (PAW) method [43], as implemented in the Vienna Ab-initio Simulation Package (VASP) [44], is used. A large plane-wave cut-off energy of 400 eV is introduced. To study the surface electronic structure, we adopt a slab-supercell approach with a slab of 9 unit cells thick and a vacuum space of 20 Å along the surface normal (001) direction. In the charge self-consistent calculations, a $20 \times 20 \times 9$ ($20 \times 20 \times 1$) k-point mesh is utilized for the Brillouin zone integration for the bulk (supercell) calculations.

### A2. Slab calculation of orbital decomposed band structure from surface atoms on S-surface

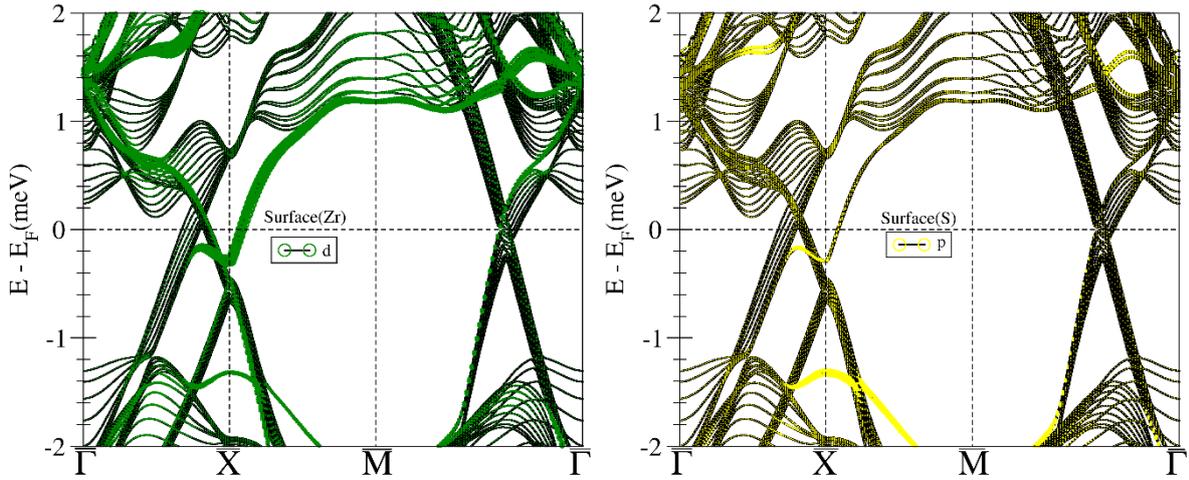

## A3. The determination of surface terminations based on step height analysis

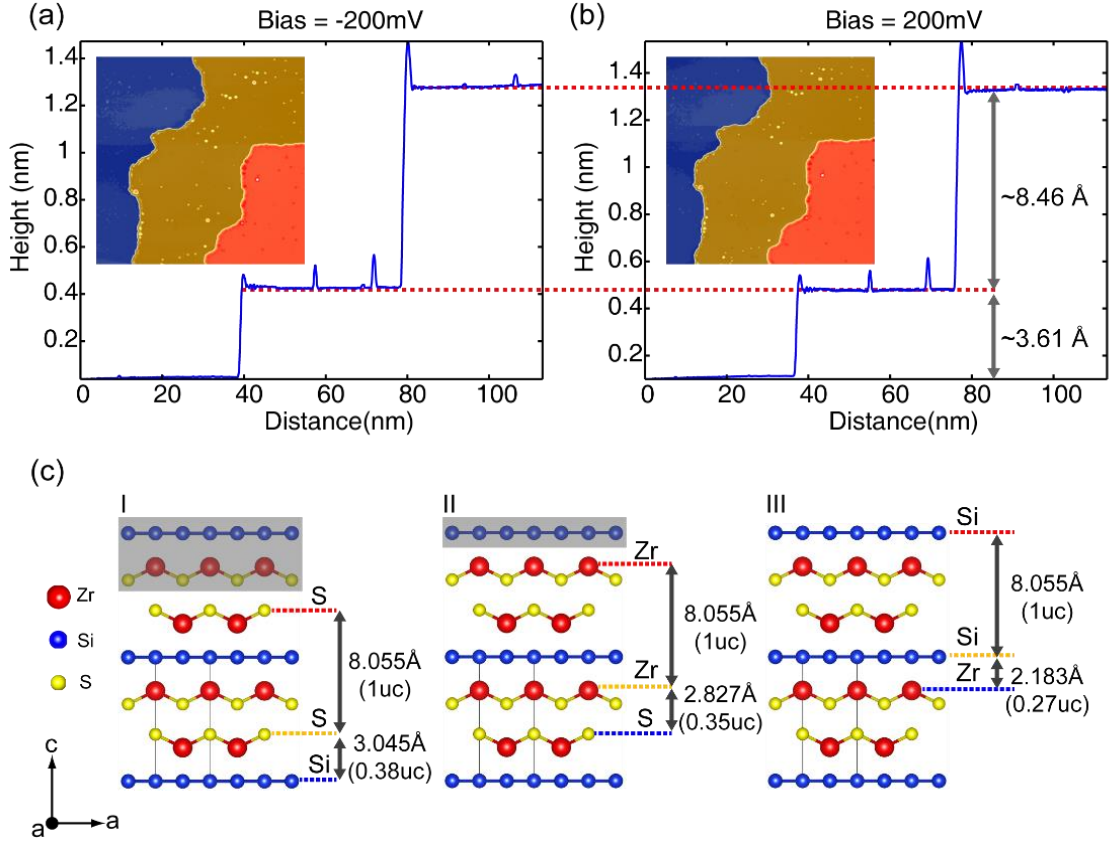

Topographic images taken in the same FOV and the lines profile along the diagonal direction with (a) bias = -200 mV and (b) bias = +200 mV reveal the identical step heights. Topographic images taken on the yellow and the red terrace exhibit identical bias dependence as shown in figures 2(b) and (d). The measured step height between them is 8.46 Å, which is approximately to 1 unit-cell (uc) of the bulk lattice. The measured step height between the yellow and the blue terrace is 3.61 Å (or 0.43 uc). (c) Three possible scenarios for the cleaved surfaces observed in Fig. 2(a). The grey regions represent the atoms being removed after the cleavage. From the comparison between the measured step height and these scenarios, we deduce the observed terraces in figure 2 agree with the Case I, which is also consistent with the STM simulation of the S-surface [17]. The Zr-termination may also exist on the cleaved surface since it is on the opposite side of our cleave plane; however, it is absent within our scan range. We note that the bond lengths among Si-Zr tetrahedron, Zr-S within quasi-2D square, and Zr-S across the gap of quintuple layers are 2.812 Å, 2.650 Å, and 2.827 Å, respectively. Thus, the observation of cleaved Si-surface is not surprising from the same level of weaker bond strength reflected on the longer bond length. In fact, the observation of cleaved Si-surface also reflects the semimetal nature of ZrSiS that the Zr-S gap between quintuple layers are not of simple van der Waals type. In addition, the sigma bonding among SiZr₄-tetrahedron requires $sp^3$ orbital hybridization for Si, which is the key controlling factor for the non-symmorphic symmetry of ZrSiS.

**A4. Topographic images of S- and Si-surfaces**

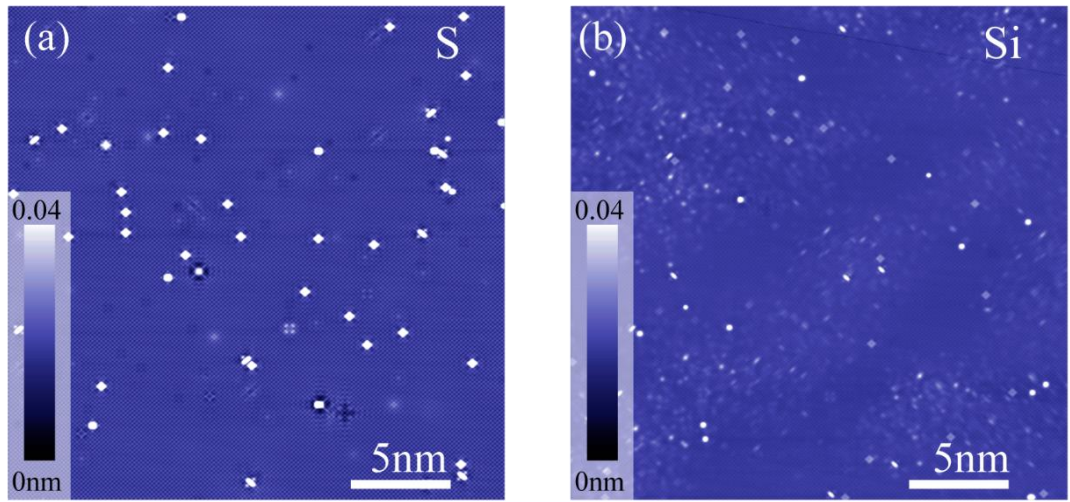

High resolution topographic images taken in the same $60 \times 60$ nm$^2$ FOV as the differential conductance maps in figure 2. (a) S-surface and (b) Si-surface.

**A5. Magnetic field dependent tunnelling spectra on the S-surface**

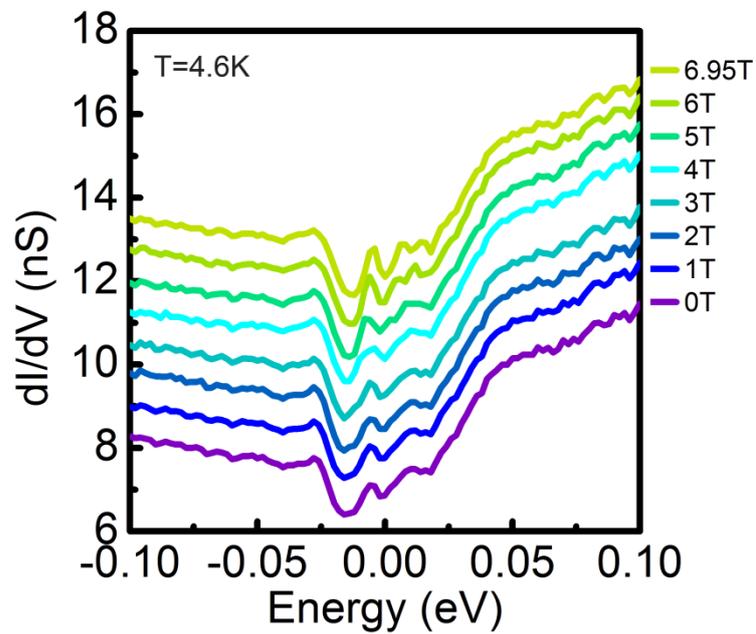